\documentclass[twocolumn,english,pre]{revtex4-1}
\usepackage[T1]{fontenc}
\usepackage[latin9]{inputenc}
\usepackage{amsmath}
\usepackage{graphicx}
\usepackage{setspace}

\makeatletter
\@ifundefined{textcolor}{}
{%
 \definecolor{BLACK}{gray}{0}
 \definecolor{WHITE}{gray}{1}
 \definecolor{RED}{rgb}{1,0,0}
 \definecolor{GREEN}{rgb}{0,1,0}
 \definecolor{BLUE}{rgb}{0,0,1}
 \definecolor{CYAN}{cmyk}{1,0,0,0}
 \definecolor{MAGENTA}{cmyk}{0,1,0,0}
 \definecolor{YELLOW}{cmyk}{0,0,1,0}
 }

\usepackage[font=small,justification=justified,format=plain]{caption}

\@ifundefined{showcaptionsetup}{}{%
 \PassOptionsToPackage{caption=false}{subfig}}
\usepackage{subfig}
\makeatother

\usepackage{babel}

\begin{document}

\title{Identifying the starting point of a spreading process in complex
networks}

\author{Cesar Henrique Comin}

\email{chcomin@gmail.com}

\author{Luciano da Fontoura Costa}

\email{ldfcosta@gmail.com}

\affiliation{Institute of Physics of São Carlos - University of S\~{a}o Paulo
Av. Trabalhador S\~{a}o Carlense 400, Caixa Postal 369, CEP 13560-970
S\~{a}o Carlos, S\~{a}o Paulo, Brazil.}
\begin{abstract}
When dealing with the dissemination of epidemics, one important question
that can be asked is the location where the contamination began. In
this paper, we analyze three spreading schemes and propose and validate
an effective methodology for the identification of the source nodes.
The method is based on the calculation of the centrality of the nodes
on the sampled network, expressed here by degree, betweenness, closeness
and eigenvector centrality. We show that the source node tends to
have the highest measurement values. The potential of the methodology
is illustrated with respect to three theoretical complex network models
as well as a real-world network, the email network of the University
Rovira i Virgili.
\end{abstract}
\maketitle

\section{introduction}

In complex network research, it is usual to study dynamics that have
a source, that is, the process taking place in the network originates
from a well-defined set of nodes, which can be sparse, appearing in
many places of the system, or clustered. There are many examples of
the latter case through the literature, including the spread of diseases
in social networks \cite{disease1,disease2,disease3}, computer virus
\cite{computer_virus1,computer_virus2,computer_virus3}, spam \cite{spam},
fads, neuronal signals \cite{nuronal1,neuronal2,neuronal3}, diseases
in a metabolic network \cite{metabolic}, and the impact of a contaminated
ambient in food webs \cite{food_web1,food_web2}, among others, so
that the study of spreading processes is one of the main topics in
this area \cite{spreading1,spreading2,spreading3,spreading4}. In
this work, we study three types of propagation: snowball (also called
dilation), diffusion, and contact process. Snowball propagation is
the classical breadth-first graph search algorithm. Although the simplest
case of the three, it can be found in the real world (e.g., a spam
network that begins with a single individual and propagates to every
contact). Diffusion dynamics in networks is closely tied with random
walks and occurs when an agent present on one node has to choose between
one of its neighbors to travel, where each neighbor has a probability
of being visited. The contact process is related to the classic disease
propagation, where each infected node has a chance to pass a disease
to its neighbors.

\begin{singlespace}
A fundamental question about a system that undergoes one of the three
processes as described above is where the origin of the spreading
is located. If this question is answered, we could, for example, know
the location where a computer virus started its contamination, the
origin of a fad, or even the origin of a disease in a metabolic network.
Little has been investigated in the literature about this matter.
Clauset and Moore \cite{bias} show that when we sample an Erd\H{o}s-Rényi
(ER) network with the snowball scheme, the resulting network has a
power-law degree distribution, which creates new topological properties
not found in the original network. Jeong \emph{et al.} \cite{revisao,stats_sampl}
made a comprehensible study on this change of properties, specially
with respect to centrality measurements, which will be the main study
in this paper. Costa \emph{et al. }\cite{trails} proposed a method
of finding the origin of trails left by agents walking through a network,
although the dilation process was performed in a different manner.
Kitsak \emph{et al.} \cite{good_spreader} studied what makes a node
a good spreader in a network, based on the \emph{k}-shell\emph{ }decomposition,
which is not a pure topological measurement and is not suitable for
our purposes of finding a single node, so we will not use it in this
paper.
\end{singlespace}

To find the source of the spreading process, we start by applying
the classical centrality measurements known as degree, betweenness,
closeness and eigenvector in the network generated by the spread.
Those measurements are discussed extensively in the works of Freeman
\cite{betw,rev_central1} and Friedkin \cite{rev_central3}, with
ideas based on the influential work of Sabidussi \cite{rev_central2}.
Then, we propose a simple modification of betweenness that accounts
for cases where the source has a very low centrality in the original
network, and show that this new measurement can provide information
about the extracted network with little influence of the original
one where the process occurred. The idea of why those measurements
should recover the source is straightforward, as the region where
the source node belongs should be central to the network generated.
So, this paper can be viewed also as an analysis of the measurements,
like the correlations that exist between them or the effectiveness
of each one. The measurements will be applied to ER and scale-free
networks in order to provide insights about the topological influence
on the success of the method, considering the homogeneity of the former
and the heterogeneity of the latter. We will also apply the method
to a real network of e-mail interchanges between members of the University
Rovira i Virgili \cite{real_network}.

The paper starts by presenting the five measurements that will be
used troughout this work. Next, we explain three methods of spreading
in networks and how they can be used to evaluate the ideas presented.
After the theoretical concepts, we show how well each measurement
performs with a snowball spreading in ER and scale-free networks,
and how the result can be improved with a simple combination of two
measurements: betweenness and degree. We then proceed to evaluate
the success of the method for the three spreading schemes. Finally,
we obtain some results based on a real network of email messages,
and show that the method is still valid.

\section{materials and methods\label{sec:materials-and-methods}}

Through this work, we use networks with two kinds of degrees distributions.
The first is the classic ER graph, which is a random graph with fixed
number of nodes \emph{N} and mean degree \emph{<k>,} where the degrees
follow a Poisson distribution. The second is the scale-free network,
which has a power-law degree distribution and can be generated following
many different methodologies. Here we consider the two most common procedures adopted
in the literature. First, one can use the Barab\'asi-Albert (BA) procedure described
in \cite{BA}, that is, starting from \emph{$m_{0}$} nodes, at each
time step we introduce a new node that makes \emph{m} new connections
with the older ones following a probability proportional to the degree
of the older nodes. The procedure is repeated many times and a network
with a power-law degree distribution $P(k)\sim k^{-3}$ is generated,
with average degree $<k>\sim2m$. Another way of constructing a scale-free
network is by using the configuration model \cite{livro_newman},
where we randomly sample \emph{N} numbers following a power-law distribution
of the kind $P(k)\sim k^{-\gamma}$ and associate these numbers with
the degree of each node, forming stubs (or half-connections) that
are randomly connected among each other with equal probability. The
networks generated by the two methodologies are a clear example of
a highly heterogeneous network because the degree distribution has
unbounded fluctuations when $N\rightarrow\infty$.

\subsection{Measuring centrality\label{sub:Measuring-centrality}}

Throughout this work, we will apply four well-known centrality measurements,
namely, degree, closeness, betweenness and eigenvector.

Let $d_{ij}$ be the length of the shortest (geodesic) path between
nodes \emph{i }and \emph{j,} then the mean geodesic distance with
respect to node \emph{i }is 

\begin{equation}
l_{i}=\frac{1}{n-1}\underset{j,j\neq i}{\sum}d_{ij},\label{eq:geodesic}\end{equation}
where \emph{n }is the number of vertices in the network. By taking
the inverse of $l_{i}$, we define the closeness centrality \cite{closeness}
of the node \emph{i}, that is,

\begin{equation}
C_{i}=\frac{1}{l_{i}}.\label{eq:closeness}\end{equation}

To define betweenness, let $n_{st}^{i}$ be the number of geodesic
paths between nodes \emph{s} and \emph{t }that passes through \emph{i},
and $n{}_{st}$ the total number of geodesic paths between \emph{s
}and \emph{t}. We define betweenness centrality \cite{livro_newman}
as

\begin{equation}
B_{i}=\underset{{s,t,s\neq t\atop s\neq i,t\neq i}}{\sum}\frac{n_{st}^{i}}{n_{st}}.\label{eq:betweenness}\end{equation}

It is usual to normalize the measurement by dividing it by $(N-1)(N-2)$,
where \emph{N} is the number of nodes of the network.

The eigenvector centrality follows the principle that a node connected
to some other high-rank node tends to have more relative importance
in the network. Let $s_{i}$ denote the score of the \emph{i}th node.
Let \emph{A} be the adjacency matrix of the network. For the \emph{i}th
node, let the centrality score be proportional to the sum of the scores
of all nodes that are connected to it. Hence

\begin{equation}
s_{i}=\frac{1}{\lambda}\sum_{j=1}^{N}A_{ij}s_{j}\label{eq:eigen_dem}\end{equation}

where $A_{ij}=1$ if node \emph{i} is connected to \emph{j ($A_{ij}=0$
}otherwise\emph{) }and \emph{$\lambda$ }is a constant. Equation \ref{eq:eigen_dem}
can be written in vector notation as

\begin{equation}
\mathbf{A}\mathbf{s}=\lambda\mathbf{s}.\label{eq:eigen}\end{equation}

The eigenvector associated with the maximal eigenvalue of this equation
represents the eigenvector centrality of the nodes.

We observe that, if we consider the usual centrality measurements,
the one that has more chances of remaining constant after the sampling
is the degree since it is a local measurement. So, we can work to
eliminate the bias caused by the original topology from which we extracted
the network by, for example, dividing betweenness by the degree of
the node. With this in mind, we define the measurement 

\begin{equation}
\hat{B}_{i}=\frac{B_{i}}{(k_{i})^{r}}\label{eq:unb_bet}\end{equation}

which is an unbiased betweenness with the proper choice of $r$. In
the results section, we will make clear why we chose betweenness
instead of closeness or eigenvector centrality (see figure \ref{fig:scatter_betw_clo}).
Also, in \cite{let_bet,let_corr} there is an interesting discussion about
the relationship of betweenness and degree on large scale-free networks.

\subsection{Spreading on complex networks\label{sub:sec_extraction}}

\begin{figure}
\subfloat[]{\includegraphics{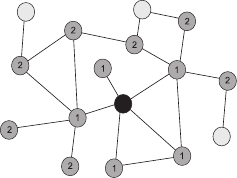}

}\subfloat[]{\includegraphics{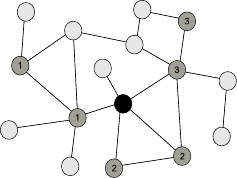}

}

\subfloat[]{\includegraphics{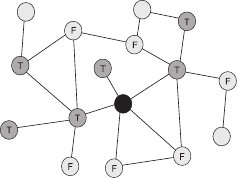}

}

\caption{\label{fig:Extra=0000E7=0000E3o_explicacao}Illustration of
the spreading schemes after two iterations, considering the black
node as seed. (a) Snowball spreading: the numbers 1 and 2 indicate
the hierarchical levels. (b) Diffusion spreading: the numbers indicate
the agent index that is executing the random walk. (c) Contact process:
\emph{T} represents nodes that accepted the contact, and \emph{F} those that did not.}
\end{figure}

Among the several types of spreading in complex networks, we focus
on those that can begin from a single node. The most common methods
used in the literature are snowball (also called dilation), random
walk, and contact process. An illustration of each method is shown in
Figure~\ref{fig:Extra=0000E7=0000E3o_explicacao}.

Snowball is a trivial spread where the subgraph is formed by the first
\emph{n} breadth-first searched nodes, forming the hierarchical levels
of a given node. Because of its triviality, it is rarely used in practical
problems, but more realistic methods tend to it on limiting cases,
so we will start our analysis by this method. In our analysis, if
the last hierarchical level can not be entirely covered, we randomly
choose nodes from it so as to achieve the desired size of the extracted
network.

On the random sample scheme, we start with \emph{R} agents inside
a unique node and let them simultaneously execute random walks through
the network, the \emph{n} first nodes visited by the agents are considered
in the final subgraph. This method reduces to snowball when we let
a large enough number of agents execute the walk. To show this, we
call $P_{i}^{h}(T)$ the probability that a node \emph{i} a geodesic
distance \emph{h} from the starting node receives a visit at iteration
\emph{T}, it is clear that

\[
P_{i}^{h}(T<h)=0,\]

\[
P_{i}^{h}(T=h)\propto1-\left(1-\frac{1}{\left\{ k\right\} }\right)^{R},\]

where $\left\{ k\right\} $ is the product of the degree $k$
of each node in a shortest path between the starting node and \emph{i}.
If \emph{R} is made large enough, $P_{i}^{h}(T=h)\rightarrow1$ and
we have again the snowball spreading. 

The contact process, well known as epidemic process in the study of
disease transmission, is done exactly like the susceptible-infected
(SI) model \cite{SI_model}, one of the simplest models of epidemics.
In the initial state, all nodes of the network are in a susceptible
state except one; then, for every connection between an infected and
a susceptible node, the susceptible node turns to infected with a
fixed probability \emph{p, }which is equal for all connections. If
$p=1$, a breadth-first transmission occurs and we have exactly the
snowball scheme explained above.

Our method consists of performing \emph{S} samplings of \emph{n} nodes
in an original network of size \emph{N}, then we apply the centrality
measurements to find each of the \emph{S }initial nodes used to start
the spreading process. Clearly, it is expected that the nodes with
the higher centrality measurements have a better chance of being the
seed, so we begin our analysis by verifying how much such measurements
separate the seed from the other nodes and how well each one performs
for the snowball spreading. Bear in mind that, from now on, spreading
and sampling have the same meaning.

\begin{figure}[h]
\subfloat[\label{fig:hist_degree}]{\includegraphics{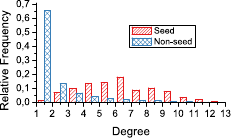}}\hspace{0.2cm}\subfloat[]{\includegraphics{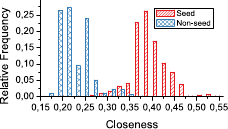}}

\subfloat[]{\includegraphics{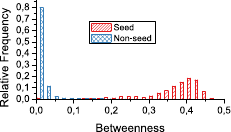}}\hspace{0.2cm}\subfloat[]{\includegraphics{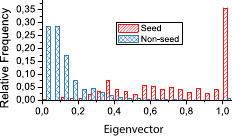}}

\caption{\label{fig:histogramas}{\small Frequency of occurrence for the four
measurements considered, separated into seed nodes (diagonal hatch)
and non-seed (checkered hatch and blue online) for 40000 nodes (400
subgraphs with}\emph{\small{} }{\small size}\emph{\small{} $n=100$}{\small ).
It is clear that, for all measurements, the seed has higher mean than
the rest of the nodes. Note that the frequency of occurrence is normalized
separately for the seed and non-seed nodes. The original network is
an ER with $N=10000$.}}
\end{figure}

\section{results and discussion\label{sec:results-and-discussion}}

\subsection{Source identification of a spreading process on artificial networks\label{sub:Source-identification-of-teo}}

\begin{figure*}
\subfloat[100 nodes]{\includegraphics{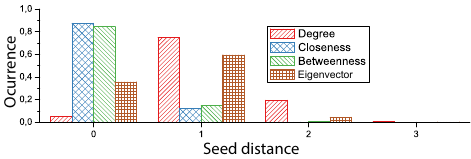}\hspace{1cm}\includegraphics{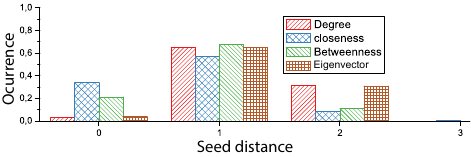}

}

\subfloat[1000 nodes]{\includegraphics{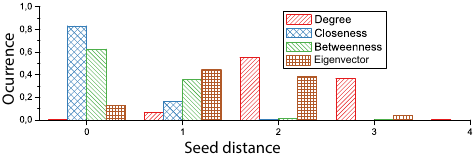}\hspace{1cm}\includegraphics{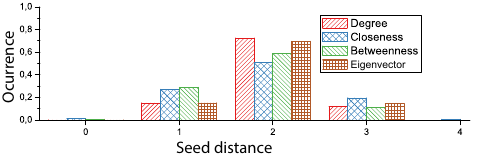}

}

\caption{\label{fig:success_rate}Histograms of the nodes with the highest centrality
measurements with respect to the distance from the seed{\small{} for
ER (left) and BA (right) networks. The parameters are the same as those used
in Figure \ref{fig:histogramas}.}}
\end{figure*}

We start our analysis with an ER network with mean degree $<k>=6$
and size $N=10000$. We sampled 400 subgraphs with size $n=100$ each.
For every subgraph, we applied the four centrality measurements discussed
above and plotted the histogram of the data separating the values
measured for the nodes used as seed and those that were not the seed.
The results are shown in Figure \ref{fig:histogramas}.

The degree distribution of the seed nodes form a Poissonian shape
\cite{bias} with mean degree $<k>\approx6$, like the original network,
but the distribution of the rest of the nodes has a scale free shape,
which is expected considering that during the extraction process we
create a small number of hubs that are close to the seed and many
low-degree nodes in the last sampled level. From the histograms in
Figure \ref{fig:histogramas}, it is clear that using the degree to
find seeds is not a good choice, while the other three measurements
have a smaller overlapping region between the two types of nodes,
especially the betweenness that appears to give the best distinction
for our purposes. 

\begin{figure}
\includegraphics{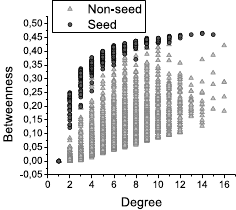}\hspace{0.3cm}\includegraphics{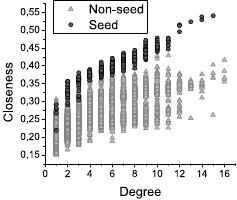}

\caption{\label{fig:scatter_betw_clo}{\small Betweenness (left) and Closeness
(right) centrality as a function of the degree of the nodes for 400 sampled networks.
The black points represent nodes used to start the sampling. The original
network follows an ER model with $N=10000$ and $<k>=6$.}}
\end{figure}

In Figure \ref{fig:success_rate} we show the number of nodes with
the highest centrality measurement divided by the number of extracted
networks as a function of the distance from the seed; clearly, the
ideal situation is when every node found has a distance zero from
the seed, which is the seed itself. We see that, in the case of the
ER model, even for a sampling of 1000 nodes, we get good results using
closeness and betweenness, which is a consequence of the homogeneity
of the network. The sampling breaks this homogeneity, as is clearly
seen by the change in the degree distribution. Considering the strong
topological bias present in the samples of the scale-free model (bear
in mind that because we randomly choose the seeds, the majority of
the samples were constructed from a very low-degree node, and such
a node usually has a hub as a first or second neighbor) the method
gives fair results for small extractions, but completely fails for
larger ones.

\begin{figure}
\includegraphics[width=0.8\columnwidth]{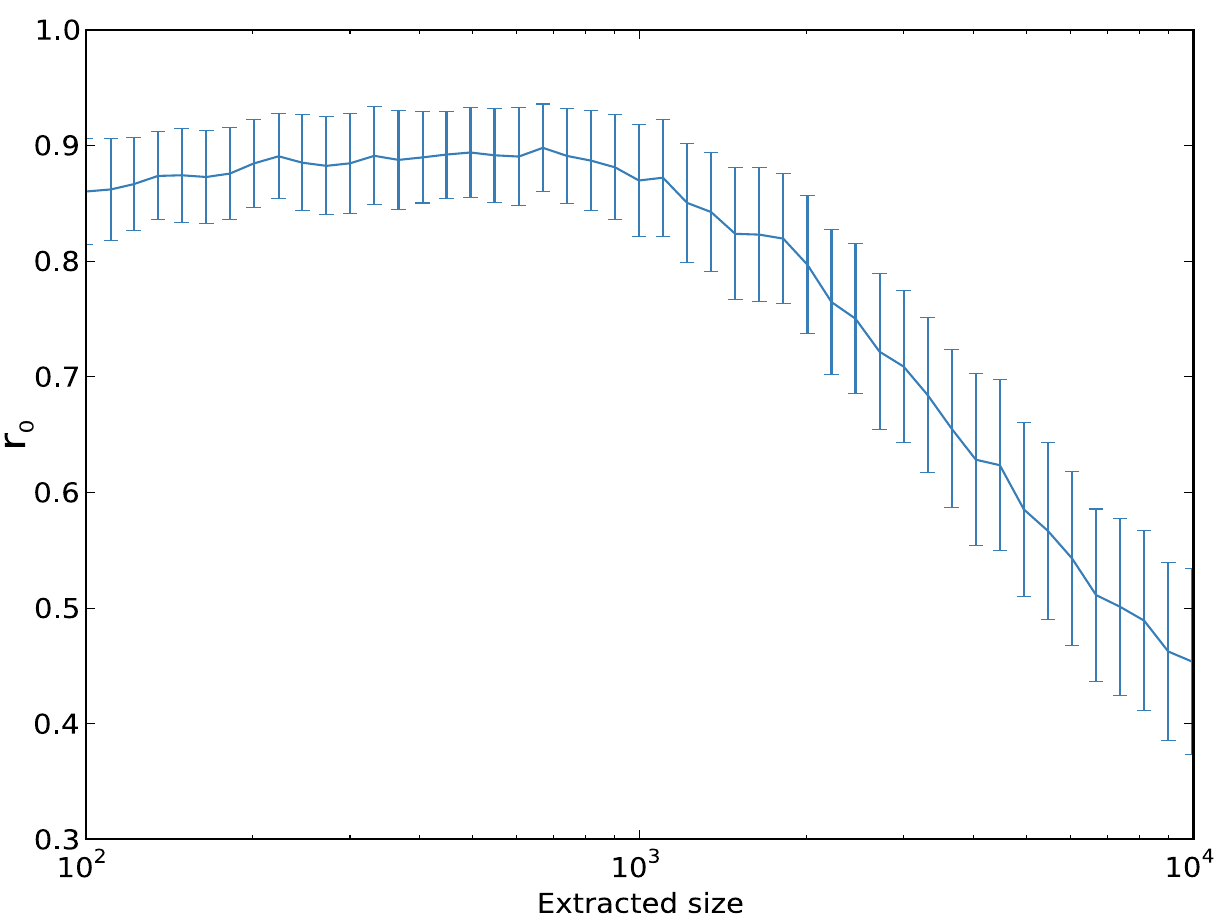}

\caption{\label{fig:fit}{\small Angular coefficients of the linear regression
between log(betweenness) and log(degree), for networks
extracted with different sizes. The original network was constructed
using the BA model with $N=1000000$ nodes and $<k>=6$. For a given
size, the mean was taken using 100 randomly selected seeds that originated
a snowball sample. The vertical bars indicate the standard deviation.}}
\end{figure}

\begin{figure}
\subfloat[\label{fig:BA_snow}]{\includegraphics[width=0.5\columnwidth]{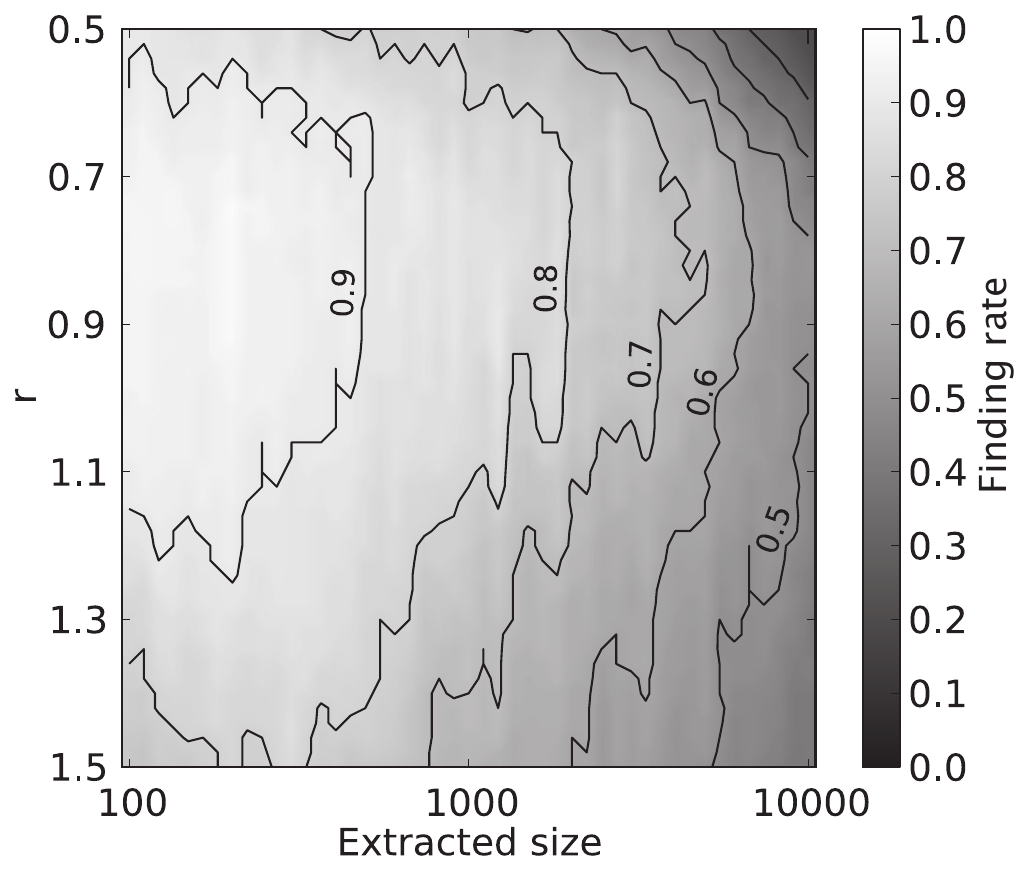}

}\subfloat[\label{fig:config_3_snow}]{\includegraphics[width=0.5\columnwidth]{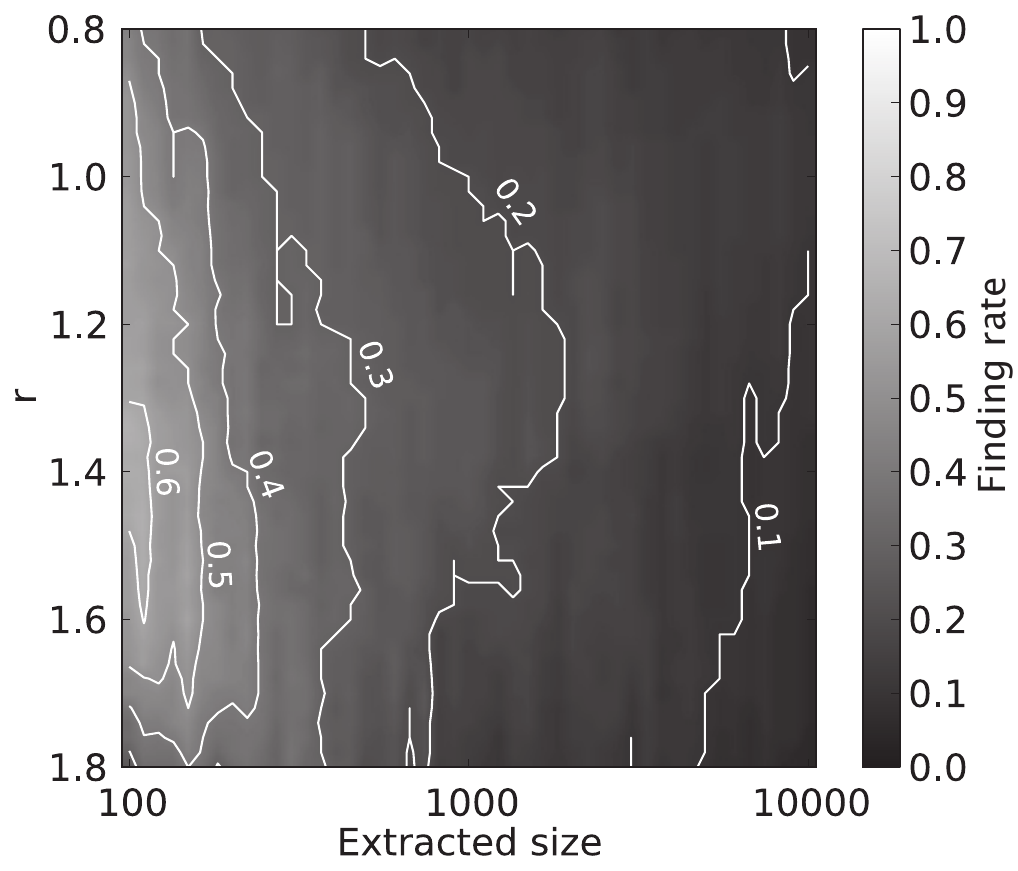}}

\caption{\label{fig:SF_model}{\small Finding rate, represented in grayscale
with contours as visual aid, of the seed as a function of extracted
size and parameter }\emph{\small r. }{\small The original network
was generated using (\ref{fig:BA_snow}) the BA procedure and (\ref{fig:config_3_snow})
the configuration model with $N=1000000$ and $<k>=6$. Note the different
scale of }\emph{\small r }{\small between the plots and the logarithmic
scale of the extracted size.}}
\end{figure}

In order to improve the results, we refer to Figure \ref{fig:scatter_betw_clo},
which shows the scatter plot of closeness and betweenness as a function
of degree when considering an ER network, constructed using the same
parameters of Figure 2, with seeds in black. We can see that, for
low degree, the closeness tends to mix the two types of nodes, which
is a property of the measurement and impossible to solve, but betweenness
mixes low-degree seeds with high-degree normal nodes, a problem that
can be solved, for example, by using equation \ref{eq:unb_bet}. We
now turn our attention to the unbiased betweenness defined by this
equation, especially to the proper value of $r$. Starting with a
BA network with $N=1000000$ nodes and $<k>=6$ we extracted, using
snowball sampling, networks with sizes ranging from $n=100$ up to
$n=10000$ (100 networks for each \emph{n}) and fitted, using linear
regression, the log-log plot of the relation betweenness versus degree.
The obtained angular coefficients are shown in Figure \ref{fig:fit};
we expect that those values of \emph{$r_{0}$} are the best choice
to define the unbiased betweenness at each extracted size, as it correctly
eliminates the bias caused if the seed has low degree compared to
other extracted nodes. To test this hypothesis, we plot in figure
\ref{fig:SF_model} the success of finding the seed (finding rate,
defined as number of correct guesses divided by number of networks
sampled) using the unbiased betweenness as a function of extracted
size and parameter \emph{r}. The original scale-free network was generated
using the BA model (Figure \ref{fig:BA_snow}) and the configuration
model with a power-law degree distribution with exponent $\gamma=3$
(Figure \ref{fig:config_3_snow}). It is clear that the model used
for the construction of the network is essential to the quality of
the method in such a way that even the best choice for \emph{r} was
different between models. For the BA network, the finding rate was
as high as 0.97 for some parameters, which is an almost perfect result,
and the best choice for \emph{r} is near the constant value of Figure
\ref{fig:fit}.

\begin{figure}
\subfloat[\label{fig:snow_gamma_2}]{\includegraphics[width=0.5\columnwidth]{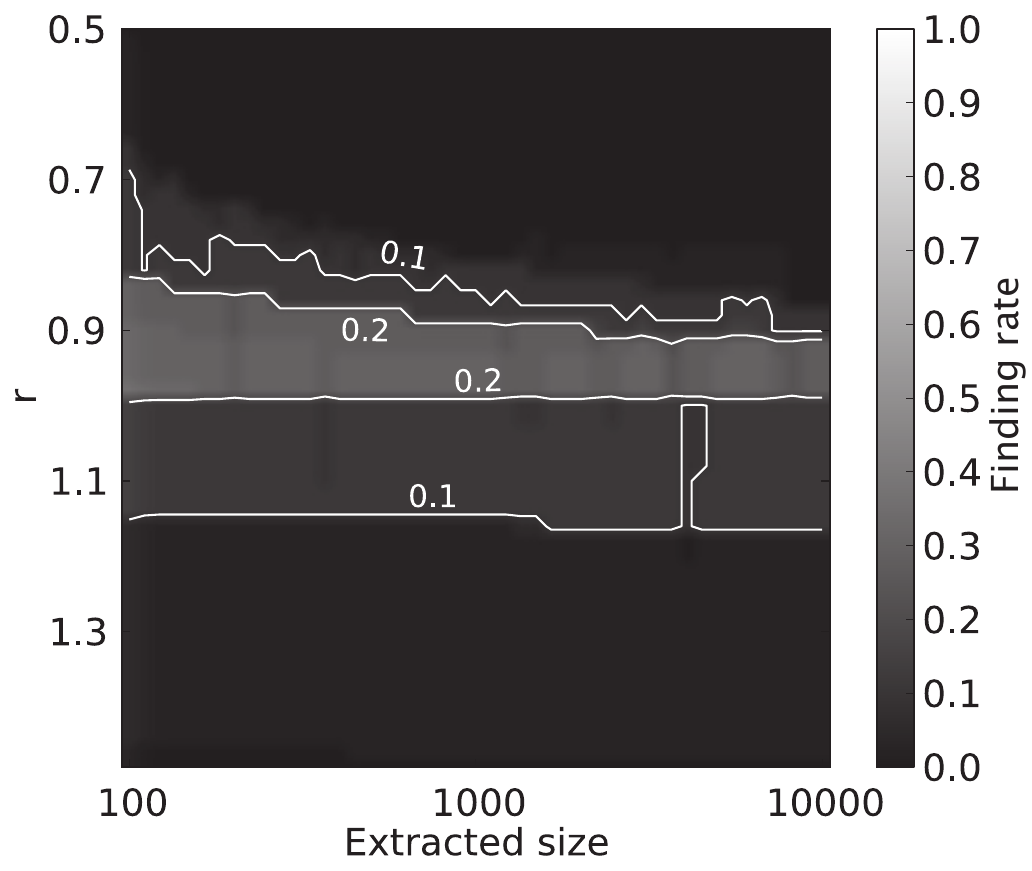}

}\subfloat[\label{fig:snow_gamma_4}]{\includegraphics[width=0.5\columnwidth]{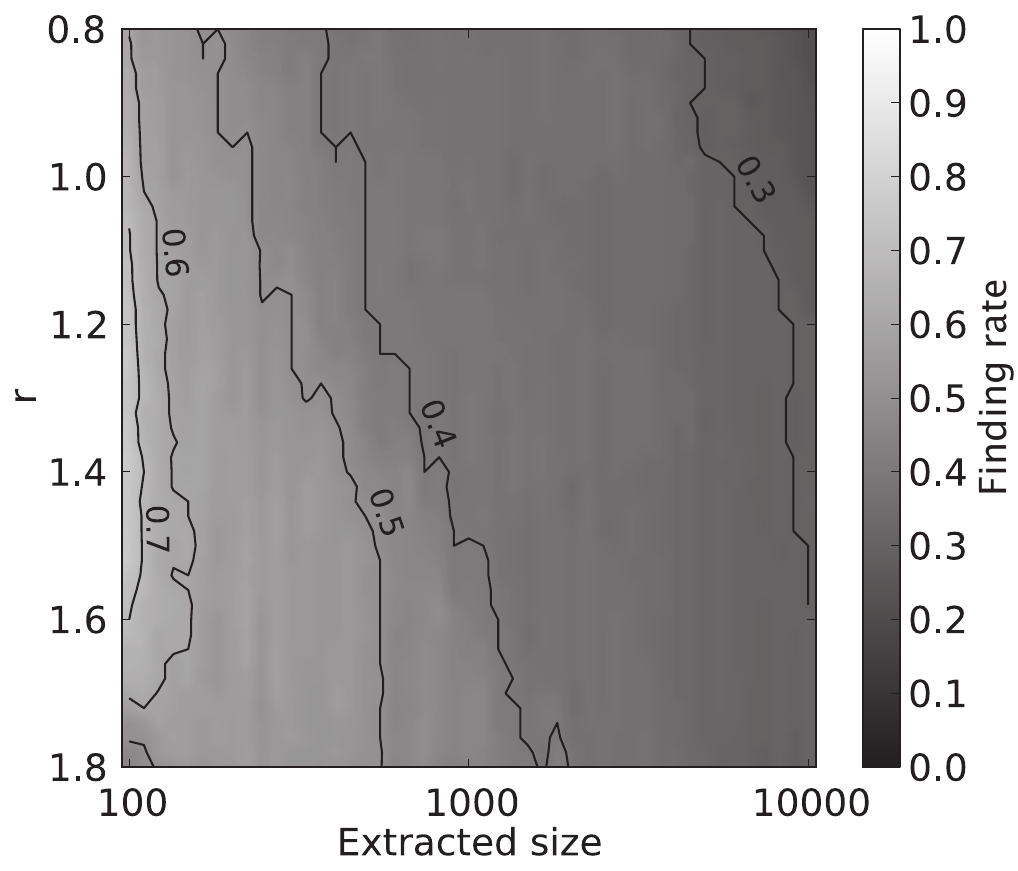}

}

\caption{\label{fig:SF_gamma}{\small Finding rate of the seed as a function
of extracted size and parameter }\emph{\small r. }{\small We used
the configuration model to generate a network with (\ref{fig:snow_gamma_2})
$\gamma=2$ and (\ref{fig:snow_gamma_4}) $\gamma=4$ with $N=1000000$
and $<k>=6$. Again, note the different scale of }\emph{\small r }{\small between
the plots and the logarithmic scale of the extracted size.}}
\end{figure}

To test the influence of the exponent of the power-law degree distribution,
the same simulation of Figure \ref{fig:SF_model} was done for other
two networks constructed using the configuration model with exponents
$\gamma=2$ and $\gamma=4$, as shown in Figure \ref{fig:SF_gamma}.
We found that larger values of $\gamma$ improve the results, while
on very heterogeneous networks (small $\gamma$), we have too many
nodes with high centrality and too strong fluctuations, leading to
a failure of the methodology.

\begin{figure}
\subfloat[\label{fig:walk}]{\includegraphics[width=0.5\columnwidth]{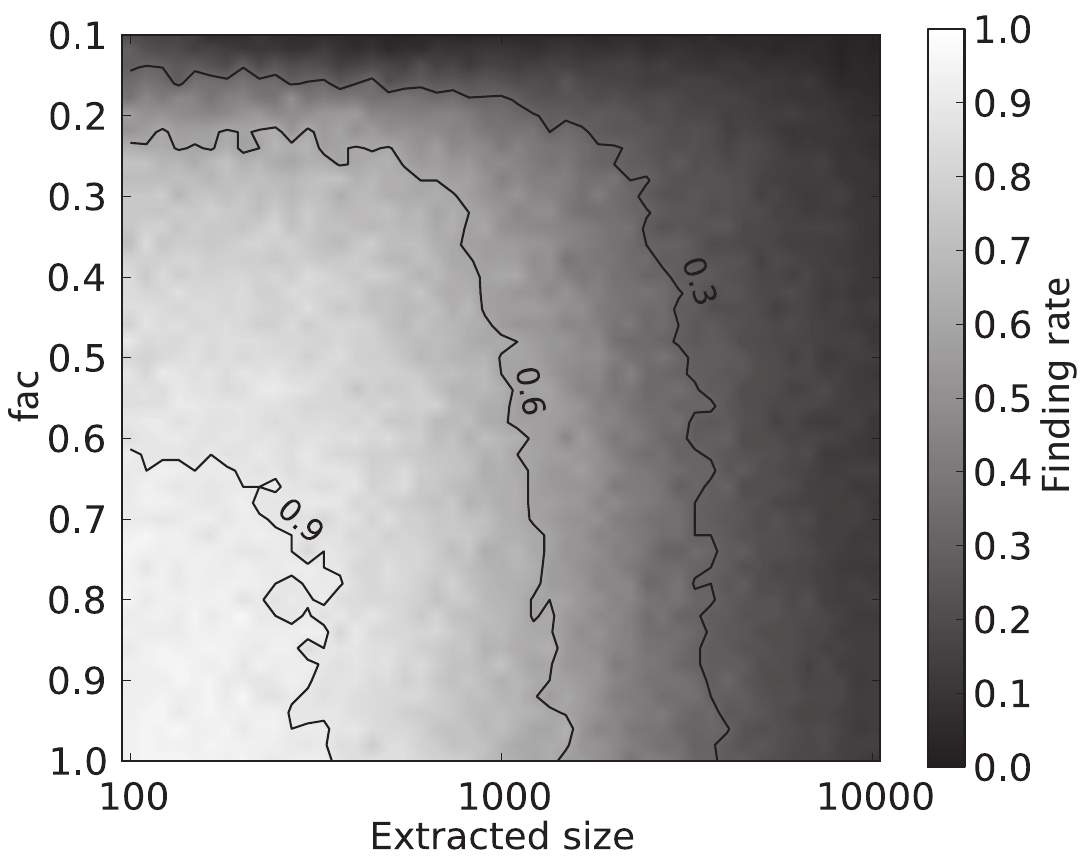}

}\subfloat[\label{fig:spread}]{\includegraphics[width=0.5\columnwidth]{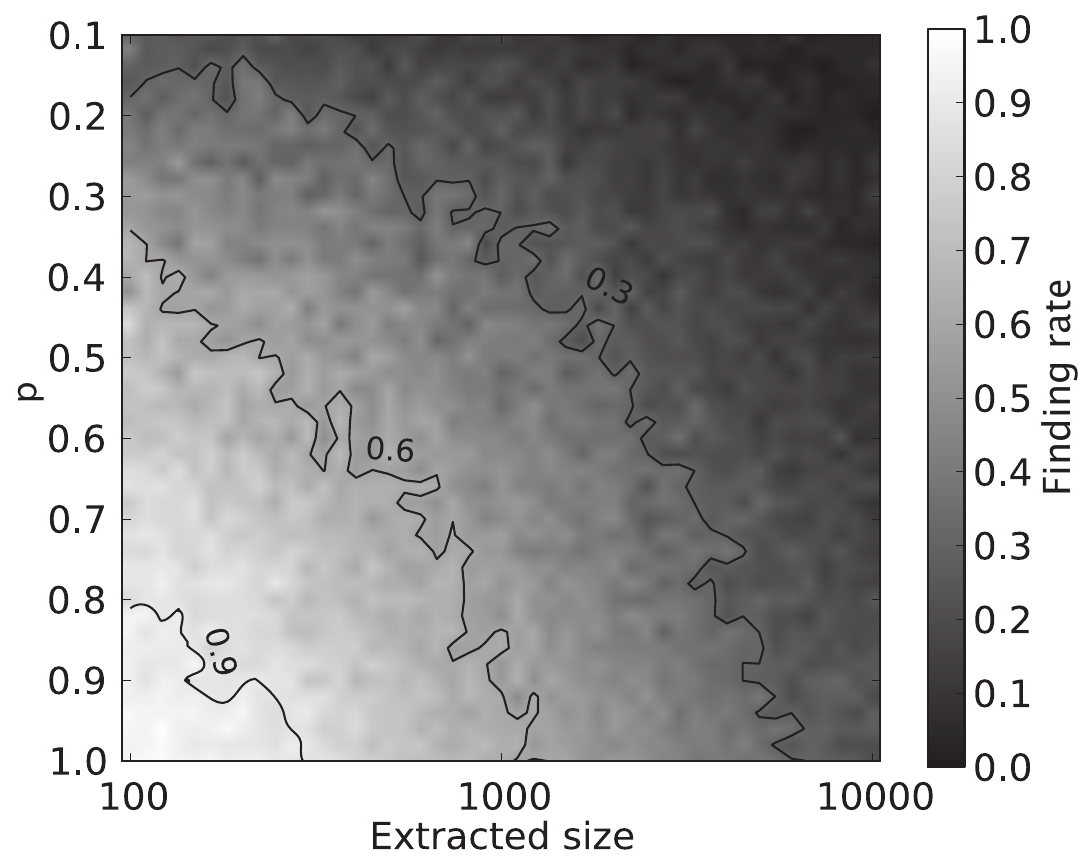}

}

\caption{\label{fig:Different-sampling}Finding rate of the seed as a function
of extracted size and (\ref{fig:walk}) number of random walkers (represented
by \emph{fac}, see text for explanation) and (\ref{fig:spread}) contagion
rate \emph{p }of the contact process. The original network has $N=1000000$
and $<k>=6$.}
\end{figure}

By using the unbiased betweenness with the empirical value $r=0.85$
suggested by the results, we can analyze the seed-finding success
for the other two spreading techniches discussed above. In Figure
\ref{fig:walk}, we show the finding rate of the seed for networks
extracted using random walkers with respect to different extracted
sizes and number of agents, represented by \emph{fac} defined as

\[
\text{number\, of\, agents}=\text{fac}\times\text{extracted\, size}\]

It is clear that, even for a small number of agents, which creates
a network composed of many chains (a sequence of connected nodes with
degree two), the method still gives fair results. We repeated the
procedure for contact processes with varying contagion rates, shown
in Figure \ref{fig:spread}, where we see that the procedure still
gives good results.

\subsection{Source identification in a real network\label{sub:Source-identification-of-real}}

\begin{figure}
\subfloat{\includegraphics[width=0.5\columnwidth]{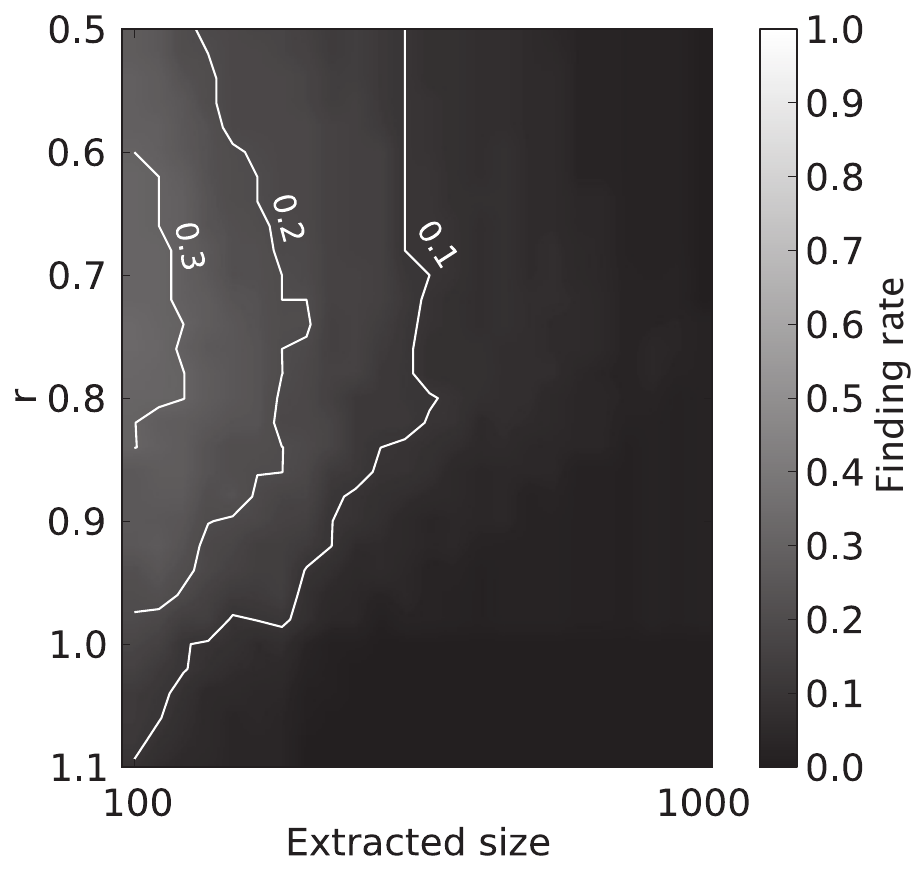}

}

\caption{\label{fig:email}{\small Finding rate of the seed as a function of
extracted size and parameter }\emph{\small r.}{\small{} The original
network represents the exchange of emails between members of the University
Rovira i Virgili.}}

\end{figure}

At last, we applied the unbiased betweenness with $r=0.85$ described
above to the email network of the members of the University Rovira
i Virgili \cite{real_network}, where each email address becomes a
node and a connection occurs if address A has sent a message to B
and B has sent an email to A. The giant component of the network contains
1133 nodes and 10902 edges, so we extracted a number of nodes ranging
from 100 to 1000\emph{ }using snowball spread beginning from randomly
selected seeds. The obtained results are shown in Figure \ref{fig:email}.
We can see that, even for a real network, it is possible to obtain
a hint about the location of the source, if the extracted (or infected)
network is small compared to the original.

\section{Conclusion}

It is clear that the identification of the seed node has great importance
for the characterization of a network originated from a spreading
process. Our purpose was to devise a method that could find this seed
node with the highest success rate possible. To do so, we utilized
four centrality measurements, which provide information about the
relative importance of a node, and applied them to diverse networks
extracted from ER and scale-free models, as well as an email network.
We found that the seed node has, in general, higher centrality than
the other nodes, so that finding the node with the highest potential
of access allows the identification of the source of the network.
When applying a single measurement, the obtained results had success
rates higher than 0.8 for large ER networks and fairly good values
for scale-free structures. We showed that a simple combination of
those measurements offers a remarkable result of more than 0.95 success
rate for small scale-free networks, considering the high heterogeneity
of the network and that the result strongly depends on the data being
analyzed, as indicated by the different results obtained between the
BA, the configuration model, and the email network. Finally, we compared
the success rate for two other spreading schemes, namely, random sample
and contact process, with varying number of walkers and contagion
rate, showing that the method works very well if the dynamic is close
enough to a snowball spreading and gives fairly good results to intermediate
parameters.

As said before, it may be possible to improve the results by combining
different centrality measurements with pattern recognition methods.
Also, we could devise a method of comparing the centrality of the
original network with that of the sampled one, which is an interesting
idea, but would require us to entirely know the original network,
which is not always possible. Finally, the method could be applied
to a network containing information about a real spreading process.
\begin{acknowledgments}
Luciano da Fontoura Costa thanks to CNPq (301303/06-1) and FAPESP
(05/00587-5) for financial support. Cesar Henrique Comin thanks CAPES
for financial support. We also thank the two anonymous reviewers for
their useful comments on earlier versions of this paper.
\end{acknowledgments}


\begin{thebibliography}{33}
\bibitem[1]{disease1}M. Small, P. Shi and C. K. Tse. IEICE Trans
Fundam Electron Commun Comput Sci, E87-A(9):2379\textendash{}2386,
(2004).

\bibitem[2]{disease2}F. Liljeros, C. R. Edling, H. E. Stanley, Y.
Aberg and L. A. N. Amaral. Nature 411, 907-908 (2001).

\bibitem[3]{disease3}J. H. Jones and M. S. Handcock. Nature 423,
605-606 (2003).

\bibitem[4]{computer_virus1}J. O. Kephartand and S. R. White. IEEE
Computer Society Symposium on Research in Security and Privacy, pages
343\textendash{}359, (1991).

\bibitem[5]{computer_virus2}R. Albert, H. Jeong and A.-L. Barabási.
Nature 406, 378\textendash{}382 (2000).

\bibitem[6]{computer_virus3}P. Holme, B. J. Kim, C. N. Yoon and S.
K. Han. Physical Review E, 65(5):056109 (2002).

\bibitem[7]{spam}J. S. Kong, P. O. Boykin, B. A. Rezaei, N. Sarshar
and V. P. Roychowdhury. arXiv:physics/0504026 (2005).

\bibitem[8]{nuronal1}L. da F. Costa and M. S. Barbosa. European Physical
Journal B, 42:573\textendash{}580 (2004).

\bibitem[9]{neuronal2}L. da F. Costa. Arxiv preprint arXiv:0802.0421
(2008).

\bibitem[10]{neuronal3}L. da F. Costa. arxiv:q-bio.MN/0503041 (2005).

\bibitem[11]{metabolic}J. S. Edwards and B. O. Palsson. PNAS vol.
97 no. 10, 5528-5533 (2000).

\bibitem[12]{food_web1}P. R. Guimarães Jr, M. A. de Menezes, R. W.
Baird, D. Lusseau, P. Guimarães and S. F. dos Reis. Physical Review
E, 76(4):42901 (2007).

\bibitem[13]{food_web2}J. Camacho, R. Guimerà and L. A. N. Amaral.
Physical Review Letters, 88(22):228102 (2002).

\bibitem[14]{spreading1}M. Boguñá and R. Pastor-Satorras. Physical
Review E, 66(4):047104\textendash{}047107,2002.

\bibitem[15]{spreading2}M. E. J. Newman. Physical Review E, 66(1):016128\textendash{}016139
(2002).

\bibitem[16]{spreading3}Y. Moreno, R. Pastor-Satorras and A. Vespignani.
The European Physical Journal B, 26(4):521\textendash{}529 (2002).

\bibitem[17]{spreading4}R. Pastor-Satorras and A. Vespignani. Physical
Review Letters, 86:3200 (2002).

\bibitem[18]{bias}A. Clauset, C. Moore, Phys. Rev. Lett. 94, 018701
(2005).

\bibitem[19]{revisao}P.-J. Kim and H. Jeong, The European Physical
Journal B, vol. 55 no. 1, 109-114 (2007).

\bibitem[20]{stats_sampl}S.H. Lee, P.-J. Kim, H. Jeong, Phys. Rev.
E 73, 016102 (2006) .

\bibitem[21]{trails}L. da F. Costa, F. A. Rodrigues and G. Travieso,
Phys. Rev. E 76, 046106 (2007).

\bibitem[22]{good_spreader}M. Kitsak, L. K. Gallos, S. Havlin, F.
Liljeros, L. Muchnik, H. E. Stanley and H. A. Makse, Nature Physics
Volume: 6, Pages: 888\textendash{}893 (2010).

\bibitem[23]{betw}L. C. Freeman. Sociometry 40:35-41 (1997). 

\bibitem[24]{rev_central1}L. C. Freeman. Social Networks, 1 215-239
(1978/79). 

\bibitem[25]{rev_central3}N. E. Friedkin, AJS, Vol. 96, N. 6 (1991).

\bibitem[26]{rev_central2}G. Sabidussi, Psychometrika, Vol. 31, N.
4 (1966).

\bibitem[27]{real_network}R. Guimera, L. Danon, A. Diaz-Guilera,
F. Giralt and A. Arenas, Phys. Rev. E , vol. 68, 065103(R), (2003).

\bibitem[28]{BA}Barabási A.-L. and Albert R., Science, 286 (1999)
509.

\bibitem[29]{livro_newman} M. Newman, \emph{Networks: An Introduction},
Oxford University Press, USA (May 20, 2010). 

\bibitem[30]{closeness}M. A. Beauchamp. Behavioral Science 10:161-163
(1965). 

\bibitem[31]{let_bet}K.-I. Goh, B. Kahng, and D. Kim, Phys. Rev.
Lett. 87, 278701 (2001).

\bibitem[32]{let_corr}M. Barthélemy, Phys. Rev. Lett. 91, 189803
(2003).

\bibitem[33]{SI_model}C. Moore and M. E. J. Newman, Phys. Rev. E
61, 5678 (2000).
\end{thebibliography}
\end{document}